# Reform-Oriented Teaching of Introductory Statistics in the Health, Social and Behavioral Sciences – Historical Context and Rationale

Rossi A. Hassad

*Abstract*—There is widespread emphasis on reform in the teaching of introductory statistics at the college level. Underpinning this reform is a consensus among educators and practitioners that traditional curricular materials and pedagogical strategies have not been effective in promoting statistical literacy, a competency that is becoming increasingly necessary for effective decision-making and evidence-based practice. This paper explains the historical context of, and rationale for reform-oriented teaching of introductory statistics (at the college level) in the health, social and behavioral sciences (evidence-based disciplines). A firm understanding and appreciation of the basis for change in pedagogical approach is important, in order to facilitate commitment to reform, consensus building on appropriate strategies, and adoption and maintenance of best practices. In essence, reform-oriented pedagogy, in this context, is a function of the interaction among content, pedagogy, technology, and assessment. The challenge is to create an appropriate balance among these domains.

*Keywords*—Reform-oriented, reform, introductory statistics, health, behavioral sciences, evidence-based, psychology, teaching, learning.

## I. INTRODUCTION

THERE is widespread emphasis on reform in the teaching of introductory statistics at the college level [1], [2]. Underpinning this reform is a consensus among educators and practitioners that traditional curricular materials and pedagogical strategies have not been effective in promoting statistical literacy [3]–[5], a skill that is becoming increasingly necessary for advancement and effective decision-making in many aspects of everyday life [6]–[8]. Statistical literacy is especially relevant given the current global information explosion [9], [10], and the emphasis on evidence-based practice and decision-making [11].

The fundamental benefit of statistics, "is that it deals with variability and uncertainty which is everywhere" [12]. Accordingly, statistics is an important tool for analyzing the "uncertainties and complexities of life and society" [13], and a "catalyst for investigation and discovery" [14]. No where is there a greater need to measure and explain variability than in the health, social and behavioral sciences, which issues directly influence and inform quality of life, a top priority in any society.

## II. OBJECTIVE AND RATIONALE

The purpose of this paper is to explain the historical context of, and rationale for reform-oriented teaching of introductory statistics (at the college level) in the health, social and behavioral sciences. A firm understanding and appreciation of the basis for change in pedagogical approach is important, in order to facilitate commitment to reform, consensus building on appropriate strategies, and adoption and maintenance of best practices. The remaining sections of this paper will address the following:

(1) A historical overview of statistics education
(2) The importance of focusing on introductory statistics
(3) The fundamentals of reform-oriented pedagogy and statistical literacy
(4) Statistics as a core course in the health, social and behavioral sciences
(5) Statistics and evidence-based practice

## III. A BRIEF HISTORICAL PERSPECTIVE OF STATISTICS EDUCATION

The practice of statistics dates back to around the mid-eighteen century, when it was regarded as "political arithmetic", given its initial sole focus on population and economic data. The practice expanded, and evolved into a scientific and independent discipline, the formalization of which can be attributed to the Royal Statistical Society (RSS, founded in 1834), and subsequently the American Statistical Association (ASA, founded in 1839). Another organization, the International Statistical Institute (ISI) was established in 1885, and it was the founding of the Committee on Statistical Education within the ISI in 1948 that initiated serious and focused dialogue on the training needs of the discipline, and research in statistics education [15], [16]. The ISI-Committee on Statistical Education (and its successor, the International Association of Statistical Education – IASE, established in 1991) emerged as the leader in this regard, with a broad

Manuscript received March 30, 2008. Rossi A. Hassad, is a member of the Faculty of the Division of Social & Behavioral Sciences, Mercy College, New York (e-mail:Rhassad@mercy.edu).

International Journal of Social Sciences 4:2 2009international focus, whereas the ASA took the charge in the USA.

For over a decade, the academic community, primarily in the USA, has witnessed reform in undergraduate statistics education, coordinated mainly by the ASA, the Mathematical Association of America (MAA), and the National Science Foundation (NSF). Specifically, there is a well-defined movement focused on reform in the teaching and learning of introductory statistics at the college level [17]–[19]. In particular, the ASA Sections on Teaching Statistics in the Health Sciences, and Statistical Education are directly involved in reform efforts and programs. In 2003, The American Statistical Association (ASA) funded the **G**uidelines for **A**ssessment and **I**nstruction in **S**tatistics **E**ducation (GAISE) project, one component of which was the introductory college statistics course. In 2005, the ASA Board of Directors endorsed the final report (and recommendations), which now serves as the blueprint for reform-oriented teaching of introductory statistics [20].

IV. TARGETING INTRODUCTORY STATISTICS

In response to the question: "If you were to start up a statistics department at a new university, what advice would you give to the new department head?", Sir David Cox (the distinguished statistician) said, "first the importance of aiming to make the first course in statistics that students receive of especially high quality and relevance" [21].

There is a long-held consensus among statistics educators that introductory statistics should be a general education requirement, and an integral part of the post-secondary curriculum [22]–[24]. Toward this end, and amidst mounting empirical evidence of students manifesting mathematics anxiety, fear, lack of interest, frustration, and emerging from the introductory course deficient in basic statistical knowledge, and with "no useful skills" [25]–[28], the statistics reform movement was formalized [29]. Over the years, the reform movement has recommended a shift in the teaching of introductory statistics from the predominantly mathematical and theoretical approach to a more concept-based pedagogy, aimed at fostering statistical literacy [30]–[32]. Statistical literacy is used interchangeably (and most times synonymously) with statistical thinking and quantitative reasoning, however, there is some debate as to the correctness of this. Nonetheless, logically, statistical literacy can be viewed as a product of statistical thinking and quantitative reasoning.

Introductory courses in any discipline are intended to provide students with exposure to the fundamentals of the field, and serve as a basis for pursuing advanced courses in that or related fields. Such introductory courses can influence students' beliefs about, and attitudes toward the discipline, and hence, to a large extent, determine whether they choose to pursue the field or go beyond the first course [33]–[35]. Moreover, for the majority of college students, the introductory statistics course will be their only formal exposure to statistics [36]. Therefore, according to Macnaughton [37], "rather than being a worst course, and possibly irrelevant, the introductory statistics course ought to be a friendly introduction to the simplicity, beauty, and truth of the scientific method".

Indeed, this recommendation reflects the spirit of reform-based teaching of introductory statistics, which is intended to facilitate statistical literacy, thinking, and quantitative reasoning, through active learning strategies, by emphasizing concepts and their applications rather than calculations, procedures and formulae. In support of reform-oriented pedagogy, David Moore [38], a former president of the American Statistical Association, wrote: "I feel strongly, for example, that statistics is not a subfield of mathematics, and that in consequence, beginning instruction that is primarily mathematical, or even structured according to an underlying mathematical theory, is misguided."

V. REFORM-ORIENTED PEDAGOGY AND STATISTICAL LITERACY

In general, the primary objective of reform-oriented teaching and learning of introductory statistics is to facilitate students to become informed consumers of statistical information [39] by addressing conceptual issues about data, such as distribution, center, spread, association, uncertainty, randomness and sampling [40]. The reform-oriented (concept-based or constructivist) approach [41], [42] to teaching introductory statistics is generally defined and operationalized as a set of related strategies intended to promote statistical literacy by emphasizing concepts and their applications rather than calculations, procedures and formulae. It involves active learning strategies, such as projects, group discussions, data collection, hands-on computer data analysis, critiquing of research articles, report writing, oral presentations, and the use of real-life data. Statistical literacy is the intended outcome of reform-oriented teaching and learning strategies, and refers to the ability to understand, critically evaluate, and use statistical information and data-based arguments [43], [44]. In essence, reform-oriented pedagogy, in this context, is a function of the interaction among content, pedagogy, technology, and assessment.

Moore [45] posited that the key to achieving statistical literacy is to facilitate students to recognize and appreciate the omnipresence of variation, and understand how such variation is quantified and explained. And Chance [46] noted that importance must be given to the context from which the data emerged, and to which the findings will be applied. These suggestions are germane to reform-based pedagogy, which characterizes introductory statistics as applied and research-oriented. Iversen [47] captures this focus in the following: "The goal of applied statistics is to help students to form, and think critically about, arguments involving statistics. This construction places statistics further from mathematics and nearer the philosophy of science, critical thinking, practical reasoning and applied epistemology."

Consistent with this approach, Hogg [48] suggested that statistics at the introductory level should be promoted as a tool of research by addressing the formulation of appropriate questions, effective data collection, interpretation, summarization, and presentation of data with attention to the limitations of statistical inference. Hogg [49] further observed

133

that "good statistics is not equated with mathematical rigor or purity, but is more closely associated with careful thinking". These notions underpin reform-oriented teaching, the core focus of which is to create active learning environments [50] which address real-world problems with real-world data, facilitating learning which is deep and meaningful rather than rote and mechanical. In other words, these active learning strategies allow students to experience the material, and construct meaning [51], an approach that is premised on the theory of constructivism [52]–[54].

Instructional design based on constructivism is generally contrasted with instruction based on behaviorism, which is described as a rigid procedural approach, intended to use fixed stimuli and selected reinforcements to promote a fixed world of objective knowledge measured primarily in terms of observable behavior [55], [56]. It focuses on discrete and compartmentalized knowledge and skills rather than integration of knowledge, and conceptual understanding. The key difference between these two approaches is that behaviorism is centered around mere transmission of knowledge from the instructor to the student (passive student and a top-down approach) whereas constructivism is focused on the construction of knowledge by the student (active student and a bottom-up approach). According to Askew et al. [57] highly effective teachers possess constructivist or connectionist beliefs rather than a transmission orientation (behaviorist beliefs).

## VI. Statistics as a Core Course in the Health and Behavioral Sciences

Statistics is increasingly becoming a core requirement for most college majors, and especially in the evidence-based disciplines, in particular, psychology, it is regarded by some academics as "the single most important course in terms of admittance into graduate schools" [58]. According to Cobb and Moore [59]: "Statistics is a methodological discipline. It exists not for itself but rather to offer to other fields of study a coherent set of ideas and tools for dealing with data." In almost every discipline, the ability to critically evaluate research findings (often expressed in statistical jargon) is recognized as an essential core skill [60] especially for college students interested in becoming practitioners in the evidence-based disciplines [61]. Consequently, undergraduate students in the health, social and behavioral sciences are generally required to take an introductory statistics course as a core requirement of their degree program [62] This course may be titled statistics, applied statistics, biostatistics, data analysis, computer applications, math, or taught as a component of an epidemiology or quantitative research methods course.

The importance and relevance of statistics to the evidence-based disciplines, specifically the health sciences, is reinforced in the following [63]: "Statistics is a core course because it provides tools needed to accurately assess statistical analyses that are reported in both the mass media and scholarly publications. The ability to effectively interpret numerical and graphical statistics is necessary for advanced study in the health professions and it is essential that health care professionals demonstrate knowledge of the statistical terminology and methodologies found in the biomedical and professional literature. The formal study of statistics complements the sciences because it also requires that students learn to formulate and test hypotheses and draw appropriate conclusions."

Undergraduate programs in the health sciences and to a lesser extent the social and behavioral sciences are unique compared to most other disciplines, in that they produce graduates who serve as licensed or certified practitioners (e.g., nursing, radiologic technology, pharmacy, health education, physical therapy, medical laboratory technology, counseling, dietetics, nutrition, respiratory therapy, physician assistant studies, occupational therapy, and medical ultrasound). Also, it is not uncommon for health sciences divisions to include programs in medicine, veterinary medicine and dentistry, and which students usually take a similar statistics course as the undergraduate students in the health, social and behavioral sciences. Graduates in these disciplines are much more likely than their counterparts in other disciplines to be required to use, produce, and communicate statistical information [64], toward evidence-based decision-making and practice [65]. Moreover, for most of these students, the introductory course may be their only formal exposure to statistics, a realization that has resulted in a call for greater importance to be given to instructional methodology, as this will affect the quality of knowledge and skills they acquire [66], [67].

This implication for teaching and learning was underscored in the Boyer commission's report on educating undergraduates [68], which emphasized that our classrooms are in crisis, and what we need, are educators in every discipline, not just subject matter experts. Earlier commentaries support this observation. Specifically, Aukstakalnis and Mott [69] noted that: "The great challenge faced by educators in every discipline is to present foreign concepts to students in forms which achieve the greatest measure of clarity and understanding." And Batanero et al. [70] cautioned that "many teachers need to increase their knowledge of both the subject matter of statistics and appropriate ways to teach the subject". This is particularly necessary given that "statistics has its own substance, its own distinctive concepts and modes of reasoning" [71]. Furthermore, several researchers and educators have noted that statistical knowledge appears to necessitate rules and ideas that to many, are counterintuitive, and therefore difficult to understand [72]–[75].

## VII. Statistics and Evidence-Based Practice

The growing importance of statistics to the health, social and behavioral sciences is largely linked to the emergence of evidence-based practice (EBP) which is defined as "the conscientious, explicit, and judicious use of current best evidence in making decisions about the care of individual patients" [76]. EBP requires that practitioners are able to identify, access, and critically evaluate relevant research evidence for reliability, validity, applicability, and overall quality, toward optimum patient care. Such appraisal and use of data necessitate statistical competence [77]. Underlying and facilitating EBP is the availability of, and greater accessibility to increasingly large amounts of research data.

The proliferation of research data in the health, social and behavioral sciences can be attributed to (among other factors) the shift in the epidemiological trend, from primarily infectious diseases to predominantly chronic non-communicable diseases, such as diabetes, heart disease, and mental health illness [78], [79]. Whereas the infectious diseases have defined and established causal pathways involving specific biological agents, the chronic non-communicable diseases present the challenge of multiple causality, as reflected in the biopsychosocial model of disease [80]. That is, there is generally no definite causal agent, but implication of multiple risk factors (including physical, biological, psychological, cultural, social, behavioral, and environmental) requiring scientific research and data analysis of different designs, and varying levels of complexity.

Another stimulus for increased research activity and data production is the expanded definition of health as "a state of complete physical, mental and social well-being and not merely the absence of disease or infirmity" [81], [82]. Specifically, these developments created a greater need for, and gave more importance to biomedical, behavioral and bio-behavioral research [83], which resulted in a broad spectrum of data of varying quality and complexity, and a challenge to decision-making, in the context of patient care. In addition to the changing epidemiological trend, and the expanded definition of health, the importance of statistical literacy was heightened by the paradigm shift in patient care associated with the primary health care approach [84].

Primary health care is intended to achieve a basic level of health care for all, and emphasizes multidisciplinary team-work rather than the traditional physician-centered model. This approach places equal responsibility and accountability for patient care on all health care practitioners, and not primarily the physician. A primary implication of the team-work paradigm is that health and behavioral sciences practitioners must now be equipped with statistical knowledge and skills to enable them to critically evaluate research evidence for reliability, validity, and applicability toward effective decision-making, in the context of patient care [85]. Hence, this new model of patient care demanded change, particularly in the health and behavioral sciences curricula, with regard to facilitating students' understanding of statistics and research.

VIII. CONCLUSION AND IMPLICATIONS

If we are to facilitate reform or change in pedagogical strategies, then we must first understand and appreciate change. Indeed, change is a process with many variables and stages, however, it begins with an awareness of the need for change. This paper posits that the emphasis on evidence-based practice is the primary underpinning of reform-oriented teaching of introductory statistics in the health, social and behavioral sciences. Professionals from these disciplines are required to demonstrate statistical literacy, in order to produce, evaluate and use research data, toward effective decision-making and optimum patient care. Consequently, statistics is increasingly becoming a core requirement for college majors in these disciplines. Associated with this development, is ongoing reform in statistics education, as in general, traditional teaching strategies have failed to promote statistical literacy. In essence, reform-oriented pedagogy, in this context, is a function of the interaction among content, pedagogy, technology, and assessment. The challenge is to create an appropriate balance among these domains. There is much empirical evidence on best instructional practices, and such information should be packaged as user-friendly manuals and audio-visual media, as a guide for instructors on how to implement reform-oriented pedagogy in their classrooms.

It is worth noting that reform-based teaching and learning strategies for introductory statistics are relatively innovative, representing a shift from the traditional, predominantly mathematical focus, to an applied, research-oriented, and concept-based approach. The underlying principles may therefore be in conflict with instructors' assumptions and beliefs about teaching and learning. This is particularly so, given that a large proportion of current instructors would have come from the traditionally taught system. Hence, resistance to this approach is to be expected, and addressed based on empirical models for initiating and maintaining positive behavioral change.

Reform-oriented instructional strategies are student-centered, and are geared toward creating active learning environments, through authentic activities (real-world tasks), so that students can meaningfully experience the material and construct meaning. Research-oriented introductory statistics courses have been shown to be advantageous in this regard, in that students become engaged is designing research, collecting, analyzing, interpreting and presenting (orally and written) data with attention to context. They often find these activities relevant and interesting, and key concepts are reinforced through applications, including the use of a computer for data analysis. Also, the epidemiological model has been proposed as a practical framework for designing introductory statistics courses to achieve quantitative reasoning [86]. This is especially applicable to the evidence-based disciplines (such as health and behavioral sciences). The epidemiological model encompasses central and unifying themes of statistics such as variability, prediction, and decision-making, with reference to real-world data, as well as salient and universal issues (health). This can facilitate students to explore data, discover meaning, and achieve deep and conceptual understanding.

Finally, unless and until substantial weighting is given to curricular development and pedagogy, as is given to peer-reviewed publications and grant awards by tenure and promotion committees, national teaching reform programs may prove to be unsuccessful. The pervasive lack of recognition of curricular development and pedagogy as scholarly and scientific endeavors, in this regard, borders on being an act of duplicity that is plaguing higher education, and undermining its core mission of effective teaching and learning.

ACKNOWLEDGMENT

I would like to express my gratitude to Dr. Anthony Coxon, Dr. Edith Neumann, and Dr. Frank Gomez for their guidance and mentorship. Special thanks to the administration of Mercy College, as well as Hunter College (Department of Psychology) for the opportunity to develop and advance my pedagogical knowledge and skills.